# Direct observation of particles with energy >10 MeV/u from laser-induced fusion in ultra-dense deuterium

Leif Holmlid
Department of Chemistry and Molecular Biology, University of Gothenburg, SE-412 96 Göteborg, Sweden
Email holmlid@chem.gu.se

**Abstract**
Nuclear fusion in ultra-dense deuterium D(-1) induced by 0.2 J pulses with 5 ns pulse length ejects ions with energies in the MeV range. The ns-resolved signal to a collector can be observed directly on an oscilloscope, showing ions arriving with energies in the range 2-14 MeV $u^{-1}$ at flight times 12-100 ns, mainly protons from the fusion process and deuterons ejected by proton collisions. Electrons and photons give almost no contribution to the fast signal. The observed signal at several mA peak current corresponds to $1\times10^{13}$ particles released per laser shot and to an energy release > 1 J assuming isotropic formation and average particle energy of 3 MeV as observed. A movable slit close to the laser target gives spatial resolution of the signal generation, showing almost only fast ions from the point of laser impact and penetrating photons from the plasma outside the laser impact point. The initial photon pulse from the fusion process is observed by a photomultiplier detector on-line.

## 1. Introduction

Nuclear fusion reactions D+D have been observed by laser-initiated processes in ultra-dense deuterium called D(-1) or d(-1) (Badiei *et al.* 2010a; Andersson & Holmlid 2012b; Holmlid 2012a). These studies have employed ns pulsed lasers of the same type as used here. Recent results with ps range pulses (Olofson *et al.* 2012) show a larger fusion rate but otherwise agree with the ns pulsed results. As shown in several publications (Badiei *et al.* 2009a, 2009b, 2010b, 2010c; Andersson & Holmlid 2009, 2010; Holmlid et al. 2009; Holmlid 2011), the D-D distance in D(-1) is 2.3 pm. This gives a density of this ultra-dense material close to $10^{29}$ cm$^{-3}$ or 140 kg cm$^{-3}$. Theoretical results for the laser intensity at break-even (Slutz & Vesey 2005) and extrapolations from experimental results on fusion in D(-1) (Andersson & Holmlid 2012b) indicated the possibility that approximately 1 J laser pulses may be useful for reaching break-even. The rather low energy value of 1 J for break-even is due to the very large density of the D(-1) material. (The density $n$ is so high that the Lawson criterion $n\tau \geq 10^{22}$ m$^{-3}$ s for D+D fusion only requires a confinement time $\tau$ of 0.2 ps).

D(-1) is a quantum material (Guénault 2003) and is both superfluid (Andersson & Holmlid 2011) and superconductive (Andersson *et al.* 2012). Similar experimental results of a superconductive state are known from very high density hydrogen clusters in voids (Schottky defects) measured by SQUIDS in palladium crystals (Lipson *et al.* 2005). The close relation between these hydrogen clusters and D(-1) has been pointed out (Holmlid et al. 2009). This may give increased fusion gains from targets with such clusters (Yang *et al.* 2011). This effect was discussed as due to Bose-Einstein condensation (Miley *et al.* 2009) or involving a Casimir effect (Hora & Miley 2007). The properties of D(-1) may be due to formation of vortices in a Cooper pair electron fluid as suggested by Winterberg (2010a, b). The structure of D(-1) observed in experiments is given by chain clusters $D_{2N}$ with *N* integer, formed by D-D pairs probably rotating around the vortex (Andersson & Holmlid 2011, 2012a). To better understand the details of the processes involved in the nuclear fusion in this material, the time-of-flight signal due to charged particles ejected from D(-1) is now studied with good time resolution. The number of MeV particles formed is so large that the nanosecond-resolved time-of-flight signal at a metal collector at 64 cm distance can be observed directly on an oscilloscope. The timing is analyzed to ascertain that indeed particles with energy > 10 MeV u$^{-1}$ are observed, and other signal sources like electrons and energetic photons are shown not to give the signals here assigned to MeV particles.

## 2. Theory

Ultra-dense deuterium d(-1) or D(-1) is the lowest energy form of deuterium atoms, but above $D_2$ molecules on the energy scale. The D-D bond distance in D(-1) is approximately 2.3 pm (Badiei *et al.* 2009a, 2009b, 2010b, 2010c; Andersson & Holmlid 2009, 2010; Holmlid 2011). This corresponds to a density of $10^{29}$ cm$^{-3}$. This material is closely related to dense deuterium D(1) which has a D-D distance of 150 pm. The rapid transformation between D(1) and D(-1) is described previously (Holmlid 2012c). A recent review of Rydberg Matter gives further background concerning the properties of D(1) and D(-1) (Holmlid 2012b). An oscillation between these two forms of deuterium was observed (Badiei *et al.* 2010c) and concluded to have a typical time of transformation of less than 0.1 second while the observed period was of the order of a few seconds. The transformation between the two forms is driven by the lower energy of the material in the form D(-1) (Andersson *et al.* 2011).

The properties of D(-1) have been amply described in the literature cited above and elsewhere. This material is stable on metal surfaces and inside the catalyst used for its formation during several days (Badiei *et al.* 2010c) in a vacuum (high or medium). However, D(-1) will react with oxygen and form water if exposed to air. It is not created by the interaction of the laser with deuterium in any form which is clear from experiments using just one pulse (Andersson & Holmlid 2011), but is formed by a catalytic process. D(-1) exists often in equilibrium with other deuterium Rydberg matter phases like D(1) and D(3) (Andersson et al. 2011; Holmlid 2012b, c). The D(-1) material consists mainly of chain clusters of the form $D_{2N}$ with *N* an integer (Holmlid 2012b; Andersson & Holmlid 2012a). *N* may take values up to 40 and above. These clusters give the superfluid properties of D(-1) at room temperature (Andersson & Holmlid 2011). At elevated temperature, a phase transition is detected (to be published) and the high-temperature phase is not yet well characterized. A non-superfluid (normal) part of the D(-1) phase has also been studied, in the form of clusters $D_4$ (Holmlid 2011). D(-1) forms superfluid layers only on metal and metal oxide surfaces. On polymer surfaces, the nucleation to D(-1) is prevented (Olofson & Holmlid 2012b). On metal surfaces, the superfluid layer may be more than one monolayer thick with the clusters standing vertically to the surface, while on polymer surfaces, the layer is thinner and the clusters lie on the surface (Olofson & Holmlid 2012b). Due to the special structure of D(-1), it interacts strongly with laser pulses of low intensity and gives Coulomb explosions (CE) in the clusters. Such processes with ejection of energetic particles take place at lower laser intensity than for ordinary materials (Badiei *et al.* 2010b).

Laser-induced nuclear fusion by inertial confinement has been investigated during several decades. Useful reviews exist for example by Hora (2007) and monographs for example by Winterberg (2010c). The direct drive spherical irradiation setup using MJ laser energy giving volume ignition was studied for example by Miley et al. (2005). Due to the large density of D(-1), nuclear fusion is expected to take place relatively easily if enough energy can be deposited in the material by for example an intense laser pulse. In the theoretical studies by Winterberg (2010a, b), ignition in D(-1) is calculated to require a laser pulse with energy of the order of 1 kJ. This is a too high value to be used in ordinary laboratory studies with no special precautions against explosion and radiation. From the calculations by Slutz & Vesey (2005) one can estimate the energy required to reach break-even in fusion in a D-T mixture. The authors discuss three different formulas, and they all indicate that break-even for D+T with a density corresponding to D(-1) will be reached at or less than 1 J laser pulse energy. Normally, D+D fusion is expected to require substantially higher energy than D+T fusion. There are however several factors which can be expected to decrease the break-even energy limit for D+D in D(-1). Both keV deuterons (Andersson & Holmlid 2010) and MeV deuterons (Holmlid 2012a) are released during laser impact, due to CE and laser-initiated self-compression (Olofson & Holmlid 2012a) processes. These effects should give a lower energy limit for break-even of D+D fusion in D(-1), and it is expected that around 1 J pulse energy will be sufficient for reaching break-even.

## 3. Experimental

The layout of the experiment is shown in Fig. 1. A Nd:YAG ns-pulsed laser is focused onto a metallic target plate with a thin superfluid layer of D(-1) (Andersson & Holmlid 2011). The focusing length of the lens is 40 cm, giving a spot size of 30 μm (for a Gaussian beam) and a power density of $3\times10^{12}$ W cm$^{-2}$. The laser is used with 532 nm light at maximum 120 mJ pulse energy, 5 ns pulse length. The particles from the target reach a collector plate at a distance of 64 cm from the target. The signal from the collector is measured by a fast digital 2-channel oscilloscope (Tektronix TDS 3032, 300 MHz, risetime 1.7 ns) with no preamplifier. The 10X probe used in some measurements is a Tektronix 500 MHz probe. Digital averaging is used in all experiments. The "drop" source for producing D(-1) is described in the literature (Andersson *et al.* 2011). In the source, a potassium doped iron oxide catalyst sample (Meima & Menon 2001; Muhler *et al.* 1992) forms D(-1) from deuterium gas (99.8%) at a pressure close to $1\times10^{-5}$ mbar. This porous catalyst is a plug that fills the opening of the gas feed tube 1-2 cm above the laser target.

The experiment is performed using a slanting (45° against the vertical) laser target and a 40 cm focusing lens in the laser beam. The target faces the laser beam, with 45° angle between the laser beam and the beam to the detectors which are both in the vertical plane. The flux to the collector is taken out at 60° towards the surface normal. The collector is a stainless steel plate with diameter 53 mm and a central opening with 20 mm diameter centered behind an annular plate with central opening of 38 mm diameter. The collector is covered by 30 μm Al foil. A tube with inner diameter 38 mm approximately half-ways between the target and the collector restricts the flux from the target to the collector. The trigger is from a photodiode (BPW34) close to the laser with a 5 V supply. A capacitor is used for coupling out the fast NIM-type trigger pulse (Nuclear Instrumentation Module standard, at -0.9 V). The diode pulse risetime is 7 ns on the oscilloscope. The coaxial cable is 2.95 m long. This length is the same as the distance travelled by the laser light from the trigger diode into the vacuum chamber to the target and from there to the collector. The trigger point is at the rise of the signal from the photodiode, at a low level relative to the peak of the trigger pulse. Since the speed of the signal in the RG58 type cable is 66% of the speed of light in vacuum as given by the material in the cable, the trigger signal is delayed by 3 ns relative to the direct light beam. This delay is taken into account by shifting the zero timing point to -3 ns as seen in the figures, giving longer times for the signal from the trigger and thus lower particle energies. No preamplifiers giving further delays are employed, but the signal is always measured directly by the oscilloscope. The timing thus has an uncertainty less than 3 ns.

In some experiments, another detector with a fast plastic scintillator and a fast photomultiplier (PMT) is used to study the signal flux. It is in the same direction behind the collector as seen in Fig. 1, and at a distance of 112 cm from the laser target. This detector has been used to measure thermal distributions of protons and neutrons in previous experiments (Andersson & Holmlid 2012b). The detection probability for thermal neutrons is low. The collector can be rotated out from the beam giving free passage for the particle beam, or it can be used as a beam-flag to stop or delay the particles in the beam to this scintillator-PMT detector. The construction of the beam-flag and of the second detector is published (Holmlid 2012a, Olofson & Holmlid 2012a). In the main vacuum chamber, a steel-plate box with a narrow slit (3 mm wide) can be rotated to analyze the flux to the detectors. The slit is at 73 mm distance from the center of the target. Its most important function here is that different parts of the target can be observed by moving the plate with its slit in the flux from the target. The plate box can also be rotated out from the particle beam as shown in Fig. 1.

## 4. Results

The signal to the collector is analyzed by attaching a 24 V shielded battery in the signal path at the apparatus, with 50 Ω input at the oscilloscope. The signals obtained are shown in Fig. 2, with the battery reversed giving +24 V or -24 V on the collector. Two different positions of the analyzing slit are used, as indicated in the figure, with the slit at maximum transmitted signal or with the slit moved out of the beam, indicated as "open". We here concentrate on the rise and peak of the signals, since these are the parts where MeV particles are observed. With a negative collector, low-energy electrons either from the target or emitted from surrounding walls by ionizing photons cannot reach the collector. (High-energy electrons from the target are excluded in separate measurements, see below). Thus, the main signal with negative collector is due to positive ions, and to electrons emitted from the collector itself, as secondaries due to the ions or possibly as photoelectrons from ionizing photons reaching the collector. The secondaries due to the ions will have a time variation similar to the ion current, so these two contributions can be treated together as an (amplified) ion current. So the only fast contribution to the signal with negative collector that is not due to fast ions is the emission of electrons from the collector due to ionizing photons (x-ray and similar) from the laser-created plasma on the target in the central chamber. This type of signal should be very fast, giving a signal at very short times, of the order of the laser pulse-length.

The first signal observed in Fig. 2, both with positive and negative collector, starts at 12 ns after the start of the laser pulse or at 8 ns from the zero in Fig. 2. When the collector is positive, an electron emission from the collector by impinging photons is not possible (or only possible with a low probability). Thus, the first signal rise in Fig. 2 which is the same with positive and negative collector cannot be due to photons. Thus it is due to fast ions. When the collector is positive, electrons from the surrounding walls emitted by photons will give a negative signal which is seen in Fig. 2 to overtake the positive signal due to ions, giving the negative maximum at 20-30 ns. This delay of 20-30 ns is due to the drift time of the low-energy electrons from the walls to the collector. .

With the slit in place and negative collector in Fig. 2, the signal peak is considerably smaller than with no slit (the "open" position). The signal observed at times shorter than 15 ns (corresponding to 10 MeV u$^{-1}$ ions) does not change with slit position, indicating a fast ion signal which moves centrally to the collector. The decrease of the peak centered at 23 ns is apparently caused by the removal and delay of these slower ions by interaction with the thin foil forming the slit. This can be seen from the reversal of the signal difference between the cases with slit and without slit in the figure, showing a delay of the ion flux in the peak. On the other hand, with the slit in place and positive collector, the signal is more positive over the whole time range than without the slit, showing that this effect is caused by a blocking of some of the ionizing photons by the slit, giving less electron current from the surroundings to the collector. This point will be further described below.

The main conclusion from the times indicated in Fig. 2 is that the signal starts at 12 ns corresponding to 14 MeV u$^{-1}$, shows a peak (partly as bump in the flank of the signal) at 15 ns (10 MeV u$^{-1}$) and has the main peak at 23 ns after the trigger, thus at 4 MeV u$^{-1}$. Further experiments using two collectors at different distances in two different setups (to be submitted) confirm this interpretation as MeV ions, giving a very accurate timing of the ion peaks. To prove the conclusions about the apparent delay of ions by the slit, a case similar to that in Fig. 2 is shown in Fig. 3. The energies of the different peaks are also given in the figure. With a negative collector in panel (b), a clear delay of the ions is observed by the slit. The target material in this case is Cu, not Ni as in Fig. 2. This shows that these features observed do not depend strongly on the metal in the target.

By moving the position of the slit, the results in Fig. 4 are found. There, a negative collector is used and the slit is moved in steps to block parts of the plasma on the target for the collector. As can be seen, the main shape of the signal peaks is similar to that in Fig. 2. However, at the extreme positions of the slit, the box structure with thicker plate blocks the particle flux more heavily, giving much lower intensity and also a clear delay of the intensity of the fastest ions at 12-15 ns. With the slit in the central position, the highest intensity is observed in the 0°-+3° range and a bump at 15 ns is clearly observed. At these central angles, the fall of the peak is also more extended than at other positions of the slit. This effect is indicated by stars in the figure. It may be due to electrons released by photons reaching the collector.

It is also possible to measure a signal to the collector with no applied voltage. This type of signal is shown in Fig. 5 with a 10X probe (10 MΩ input) connected to the collector at a much higher impedance level than with direct measurement as in Fig. 4. This is probably the reason for the pick-up of disturbances from the laser prior to the actual signal rise. Oscillations are also caused by bad impedance matching at the collector. The oscillations are relatively small as shown in Fig. 5 when using the 10X oscilloscope probe with 50 Ω input to the oscilloscope. The main features from Fig. 4 are retained, with almost identical peak structure. However, the slow fall of the signal after the peak in the range 45-48º seems to have disappeared. This indicates that this part of the signal in Fig. 4 is due to electrons released from the collector by photons. Such electrons will not be able to leave the collector in this case with no voltage bias.

The signal to the outer scintillator-PMT detector shows clearly that the signal from the laser focus on the target is due to fast particles. Many experiments with this detector have been done by moving the slit in front of the target as in Figs. 4 and 5. The spectra obtained vary strongly in intensity with slit position and time-of-flight and thus a full set of results as in Figs. 4 and 5 requires several figures and is not shown here. See also similar results in Holmlid (2012a) and Olofson & Holmlid (2012a). In Fig. 6, the signal measured at a few angles is shown with the collector in place and operating as a beam-flag for fast ions. At the central positions of the slit where the collector signal is at its maximum in Figs. 4 and 5, almost no fast signal is observed with the scintillation detector. The results in Fig. 6 are from the same experiment as in Figs. 4 and 5 with unchanged conditions otherwise. At other slit positions as shown in one example in Fig. 6, a high intensity signal is observed from particles (photons) which penetrate through the beam-flag, giving almost the same intensity with the beam-flag closed and open. Such results are shown in Fig. 7. (This measurement has been done under slightly different conditions to avoid signal saturation in the PMT or its preamplifier thus retaining the time resolution). Thus, at these slit positions the main signal observed is due to penetrating photons from the fusion plasma. The lay-out of the slit relative to the target and the plasma is shown in Fig. 8. Thus, at a slit position of -4º the plasma contributes most of the signal, even if a few MeV particles pass the slit and are observed delayed at 200-500 ns in Fig. 7. Further results of the same type are given in Holmlid (2012a) and Olofson & Holmlid (2012a).

Further tests have been made to ascertain that no fast electrons are involved in the signal to the collector. A strong electric field from a -7 kV supply was used (inside the closed box with the slit) when the beam to the collector passed through the plate box, with no signal change. A few small strong permanent magnets were attached to the apparatus midways to the collector, at a distance of 2 cm from the beam, with no signal change. These tests make it certain that no high-energy electrons move from the target to the collector and influence the signal observed.

It is possible to observe even faster positive ion peaks than shown above, by using a higher laser repetition frequency of 15 Hz, or by directing the laser beam onto the catalytic emitter material in the D(-1) source. Such data are shown in Fig. 9, with the peak of the distribution at 13 ns or 14 MeV u$^{-1}$. These results indicate directly that particles with energy > 10 MeV u$^{-1}$ are produced by the fusion processes.

One further effect of the slit in its central position can also be seen in Fig. 10. There, it is shown that the slit does not decrease the observed signal to the collector which has zero bias, but instead the signal is increased with the slit in place. The main possibility for such an effect is that a counteracting signal is blocked by the slit, i.e. one that adds a negative signal which reduces the positive signal observed. This may be due to emission of electrons from the surroundings of the collector without the slit in place, for example by fast ions.

Finally, the variation of the peak measured with negative collector is shown in Fig. 11 as a function of the laser pulse energy. The size and shape of the peak, the position of its maximum and the first signal rise after the trigger are all changed. The peak height varies approximately as (energy)$^4$. The peak and the rise of the signal are both delayed by 5 ns at low pulse energy. This effect is not due to changes in the trigger signal which was recorded simultaneously. The trigger decreases in size from -1.6 V to -1.2 V at its peak, with a very small change in the peak shape. If the collector signal would be caused by photons from the laser impact through electron emission from the collector, no delay at lower pulse energy is expected. However, the number of collisions for ejected particles is likely to increase with lower pulse energy, giving both a delay and smaller signal, as observed. Thus, the results in Fig. 11 further support the conclusion that the collector signal is due to MeV particles.

## 5. Discussion

To conclude that particles with > 10 MeV u$^{-1}$ energy are observed, it is necessary to ascertain that the timing used to determine the time-of-flight is correct, and that neither photons nor electrons give the signals observed. The measurements of the timing have been described above and there are no apparent factors that would change the conclusions about the timing being accurate within 3 ns. A problem would exist if the trigger signal from the diode at the laser was delayed further than calculated on its way to the oscilloscope. In such a case, the TOF values observed would be longer than concluded from the data. However, there are no reasons to assume that. The diode is biased by a small power supply, and the trigger is ac coupled through the cable to the oscilloscope. There is no amplifier step or similar used. Thus, the timing system is as simple as possible to avoid delays of the trigger and the timing used is concluded to be accurate within 3 ns.

The appearance of the signals is also in agreement with the timing calculated. For example, the negative current observed with positive collector as in Figs. 2 and 3 arrives a few ns later than the positive current since the low-energy electrons from the surrounding parts will take some time to reach the collector, even with the +24 V driving voltage. It is also of interest to note that the initial rise of the signal in Fig. 2, up to 15 ns after the trigger, is unchanged by the slit position in Fig. 2. Thus this signal is due to fast ions not influenced by the slit, while the signal after 15 ns varies with the slit position since the slit will block or delay these slower ions. The observation of fast positive peaks at 15 ns with both signs of the collector voltage confirms strongly that fast positive ions are observed with the timing expected. Photons cannot give a positive signal at the positive collector.

The tests described in the Results section exclude any contributions to the signals from fast electrons from the target. Slow electrons, ejected by the impact of ionizing photons from the walls and other parts surrounding the collector, will contribute to the signals observed, as described for example by Figs. 2-3 and Fig. 10. However, these currents are relatively small and do not change the main features of the fast signals. The direct interaction of the photons with the collector must also be considered. The initial photon pulse from the fusion process at the target should give ejection of electrons from the collector. The electrons ejected may give a positive signal with negative collector. With positive collector, a smaller signal will be observed since the electrons will be retained by the positive voltage. However, if the slit is moved to mainly observe the photon signal with the scintillator-PMT detector as in Fig. 7, the photon pulse is easily observed. It is also seen directly from such results that most initial pulse photons pass through the collector Al foil and do not interact strongly with the foil. Thus, the contribution to the fast signals from energetic photons is small also due to this effect.

The collector signals thus show clearly that MeV particles with several MeV energy, up to 14 MeV $u^{-1}$ are ejected from the laser target. The process for this is D+D fusion and the particles ejected are protons and possibly also deuterons (accelerated through collisions with protons). It is not expected that neutrons will be observed in the collector signal due to the weak interaction with the Al material, and both channels in D+D fusion are expected to contribute to the signal thus giving both 3.02 MeV and 14.7 MeV protons. Deuterons can obtain energies up to 1.3 MeV $u^{-1}$ and 6.5 MeV $u^{-1}$ from linear collisions with such fusion generated protons. The timing of the signal shows that it corresponds to the expected initial proton energies from D+D fusion.

Since the protons from the fusion processes have to leave the ultra-dense layer of D(-1), collisions with deuterons are possible. The calculated range for the fusion generated protons in ultra-dense D(-1) is very small, 12 nm for a continuous dense phase (Holmlid 2012a). Thus, for the protons to leave the D(-1) layer, a large disturbance may be needed, like a fusion process ejecting numerous particles in a few ns time. The most likely mechanism slowing down protons in D(-1) is the reflection of protons from the ultra-dense D(-1) layer below the point of laser impact. Such a process gives slower protons after the collisions. Protons initially with 14 MeV energy moving into the D(-1) layer will return as 1.5 MeV protons with time-of-flight 39 ns, and protons initially at 3 MeV will be reflected as 0.33 MeV protons with a time-of-flight of 83 ns. Such protons are still within the observed peak width in Figs. 2 and 3. Non-linear collisions will give smaller energy losses to the deuterons. Thus, the broad peaks obtained are probably due to protons colliding with (reflecting from) deuterons during the ejection from the fusion process.

In the fusion process, also other types of particles should be formed. Since > 10 MeV $u^{-1}$ particles are observed, it is clear that the fusion process has not ended by formation of $^3$He and T together with protons and neutrons, but has continued by the reaction of $^3$He and T with D to form 14 MeV protons and neutrons, and $^4$He at 3.5 MeV. These alpha particles may also be observed. Their TOF to the collector is close to 50 ns and they may exist in the collector signal at > 50 ns due to scattering. For example in Fig. 4, the signal with the slit centered does not fall rapidly in this TOF range due to photoelectrons (see further above) and may thus contain these alpha particles which avoid direct observation. $^4$He has been detected selectively in other experiments to be published. The interaction of $^4$He with the superfluid D(-1) is of special interest in such experiments.

A discussion of the absolute signal observed must include a discussion about the secondary electron emission coefficient for the impact of MeV protons on a metal collector surface like Al used here. Such data exist in the literature (Park & Jang 2007; Borovsky *et al*. 1988; Thornton & Anno 1977). The coefficient is smaller than unity for this energy range and thus the signal observed with positive collector should be less than twice as large as the true proton signal, which is similar to the signal found with the collector at ground potential. The total charge observed at the collector is thus up to $3\times10^{-10}$ As or $2\times10^{9}$ ions per laser shot. The exposed collector area corresponds to $2\times10^{-4}$ of the full sphere, which gives a total charge ejected of $1.5\times10^{-6}$ As or $1\times10^{13}$ ions per laser shot assuming isotropic formation of the protons. That this value is considerably higher than the previous value given in Holmlid (2012a) may be due to the actual small viewing factor in these earlier PMT based measurements, which was difficult to estimate correctly. If this charge is due to an average particle energy of 3 MeV, the total energy in the ejected protons is 4.5 J per laser shot. Assuming that the ions detected are not protons but that they are heavier than protons means that the energy observed is correspondingly larger. Thus, the most conservative (lowest) estimate is that all the ions are protons. The observed peak ion current is 8 mA, which corresponds to 16 A in the peak, assuming isotropic formation.

A related point to discuss may be the nature of the fusion region. It is clear that the region where the fusion takes place is not very hot on average. In similar experiments, the neutron temperature is found to be 50-100 MK (Andersson & Holmlid 2012b; further results to be published). The laser intensity has its maximum in a small volume, of the approximate diameter of 30 μm which is the laser beam waist. In this volume, the D(-1) layer may have a thickness of the order of 1 μm. Thus, the total volume of interest is approximately $1\times10^{-9}$ cm$^3$, containing a maximum of $10^{20}$ D atoms. If the D(-1) layer is thinner, the number of deuterons is smaller, giving a larger average energy per deuteron from the laser pulse impact. (This situation is very different from the more often tested hydrogen ice case, where the density is a factor of $10^3$ - $10^6$ lower giving a much higher average energy from the laser pulse to the atoms). The number of ions ejected here is $1\times10^{13}$ per laser shot, thus only $10^{-6}$ – $10^{-7}$ of the existing deuterons participate in the fusion processes. That such a small fraction of the available deuterons is involved is mainly due to the low laser pulse-energy available and the long pulse-length of 5 ns. With a laser pulse-energy a factor of 100 larger, the fraction fusing may be large enough to create a plasma of high enough temperature to give ignition.

## 6. Conclusions

It is shown that massive particles are ejected with energies above 10 MeV u$^{-1}$ by pulsed laser impact on a layer of ultra-dense deuterium D(-1). Electron and photon contributions to the fast signals are identified and found to be small. Fusion processes have been observed previously in this material, and the present results confirm that nuclear fusion D+D is initiated by laser pulses of energy 0.2 J and 5 ns length. Direct signal collection by a metal collector at a distance of 64 cm is used. Peak currents of ions up to 8 mA are observed, corresponding to a total ejected current of 16 A assuming isotropic emission. The total energy in the protons formed is > 1 J, assuming isotropic formation.

# Figure captions

Fig. 1. Horizontal cut through the apparatus, not to scale. The box in the central part is movable and has an entrance slit 3 mm wide at a distance of 74 mm from the target while the exit slit is much wider. The distance from the target to the collector is 640 mm. The collector can be rotated out of the beam. The target-scintillator distance is 1120 mm.

Fig. 2. Collector signals with a 24 V battery in the cable to the 50 $\Omega$ oscilloscope input. The voltage of the collector is indicated. The position of the slit in the chamber is indicated as "open" (slit turned out of the beam) or "slit" at the maximum transmitted signal. Times from zero are shown. Ni target.

Fig. 3. Collector signals with a 24 V battery in the cable to the 50 $\Omega$ oscilloscope input. Panel (a) is the signal when the collector is positive and attracts negative charges, while panel (b) shows the signal when the collector is negative and repels electrons, also those ejected from the collector by impact of heavier particles. The curves with the slit at maximum signal are shown, and the unmarked curves are measured with the slit moved out from the beam. The times given are from the zero indicated. Cu target.

Fig. 4. Collector signal with -24 V on the collector and 50 $\Omega$ oscilloscope input. The angles to the right indicate the angle relative to the nominal maximum transmission position of the slit. The vertical lines indicate the start of the signal and the maxima. The arrows indicate the shift of the maximum at large angles with low transmitted signal. The stars indicate long slopes of the signals. Ni target.

Fig. 5. Signal with a 10X probe from the collector to the oscilloscope. Compare with Fig. 4. The vertical lines indicate the start of the signal and the maxima. The arrows indicate the shift of the maximum at large angles with low transmitted signal. Ni target.

Fig. 6. TOF peaks in the scintillator-PMT detector with the beam-flag (the collector) in the beam, at angles shown to give maximum transmission of the particle signal in Figs. 4 and 5. Note that almost no particle signal passes un-deflected through the collector at the central positions of the slit. Ni target.

Fig. 7. TOF peaks in the scintillator-PMT detector signal at negative angles of the slit, thus mainly observing the plasma. Beam-flag (collector) position both closed and open (out from the beam). The signal is mainly due to photons which penetrate the beam-flag. A low-intensity delayed particle signal is also observed. Al target.

Fig. 8. Principle of the selection of flux to the detectors by the movable slit (box) at the target. In the position shown, particles from the point of laser impact have to penetrate the steel slit edge giving a delayed signal. Photons from the plasma are observed by the detectors on-line.

Fig. 9. Fast peaks due to high repetition rate of the laser (upper curve) and laser hit of the catalyst in the source (lower curve). Collector signal with a 10X oscilloscope probe, no bias. Ta target.

Fig.10. Peaks with the slit at the maximum transmission, and "open" with the slit moved out of the beam. The slit removes ionizing photons and thus an electron signal from the walls to the collector. Ta target, no bias at the collector.

Fig. 11. Variation of the collector peak with laser pulse energy, as indicated. Negative collector, coaxial 50 Ω oscilloscope input. Cu target.

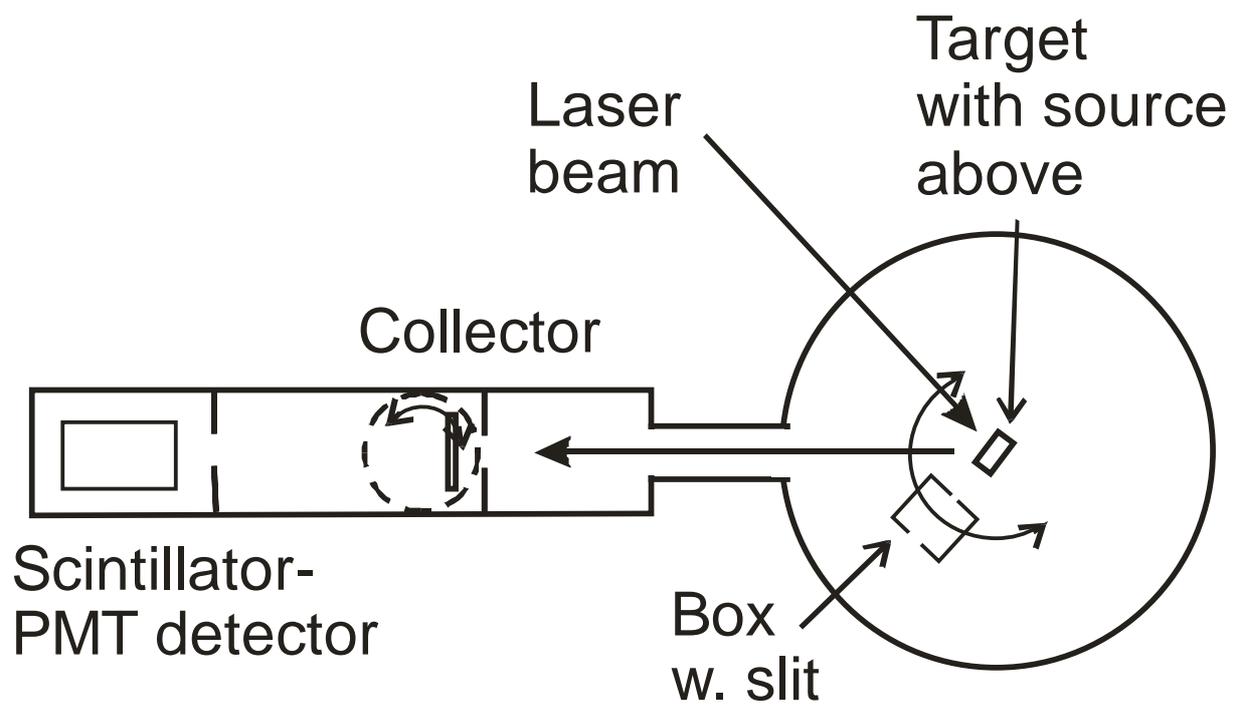

Fig. 1

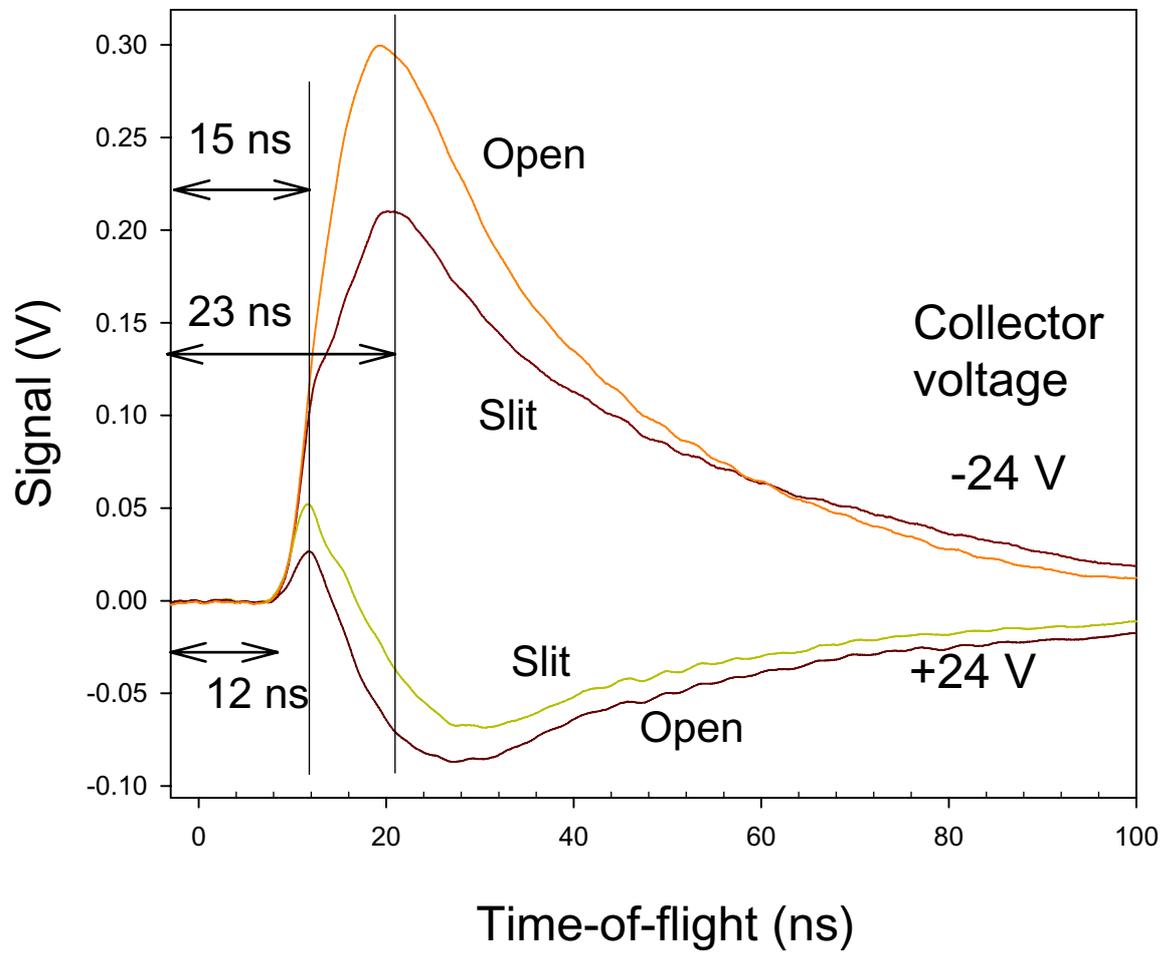

Fig. 2

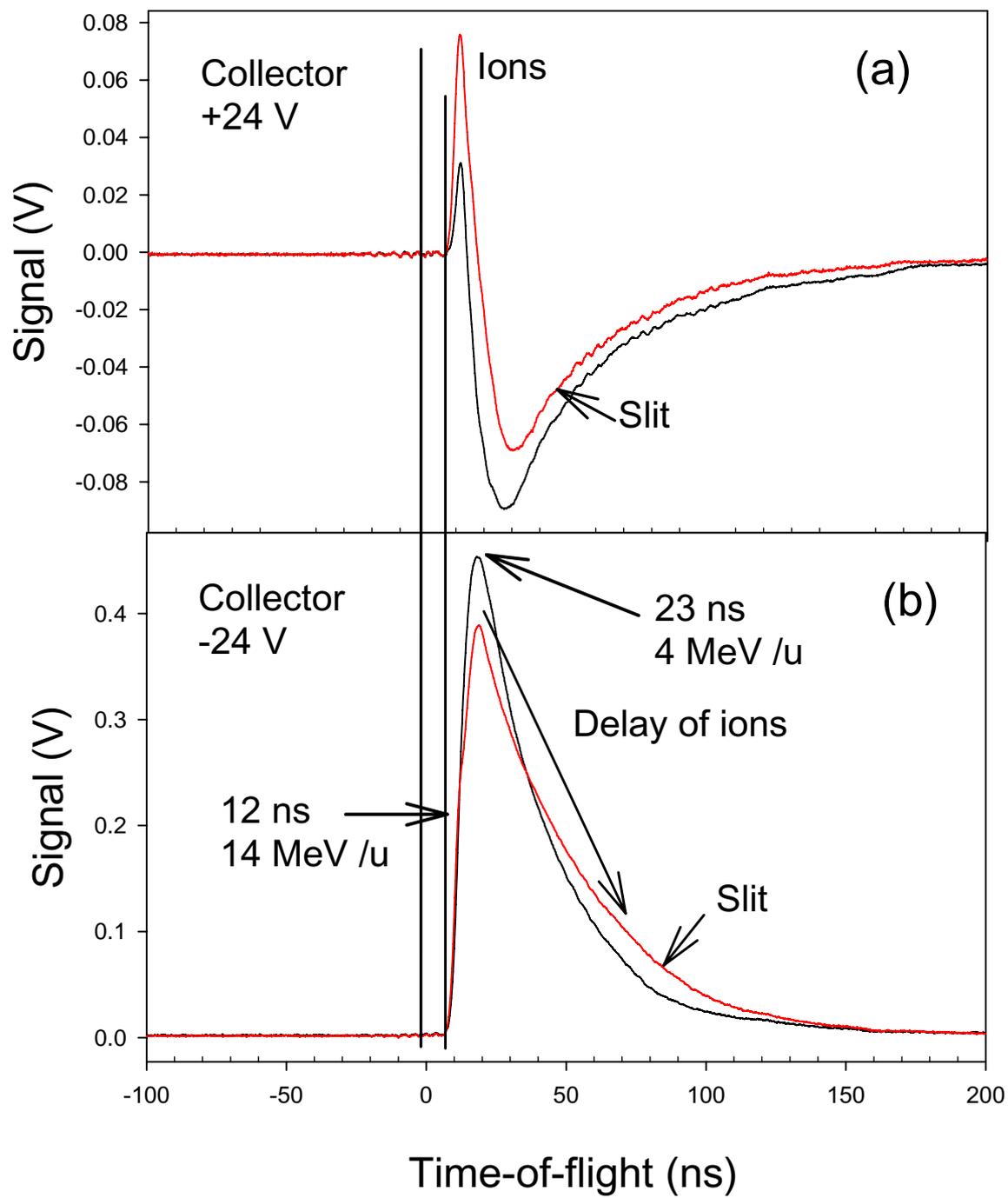

Fig. 3

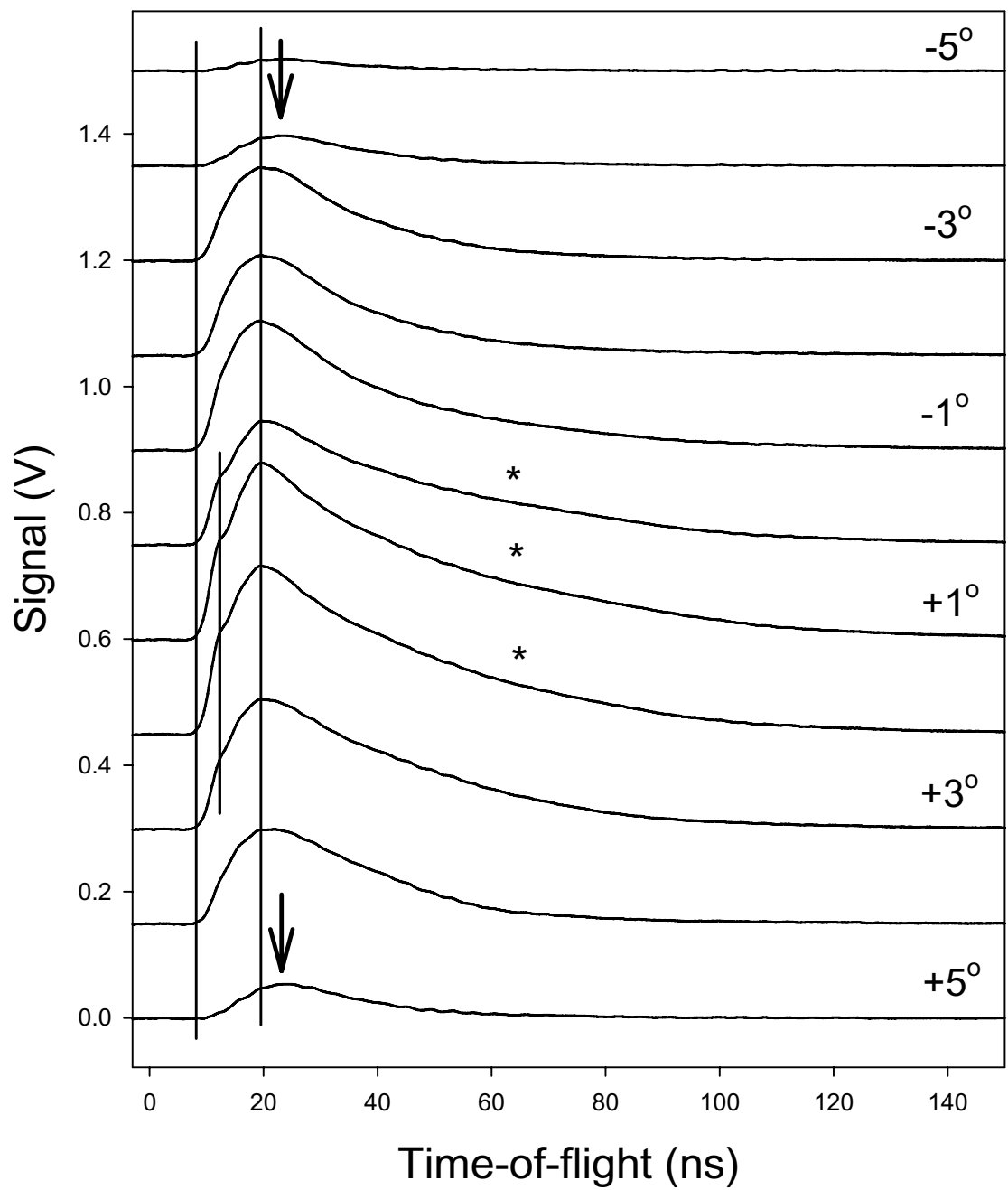

Fig. 4

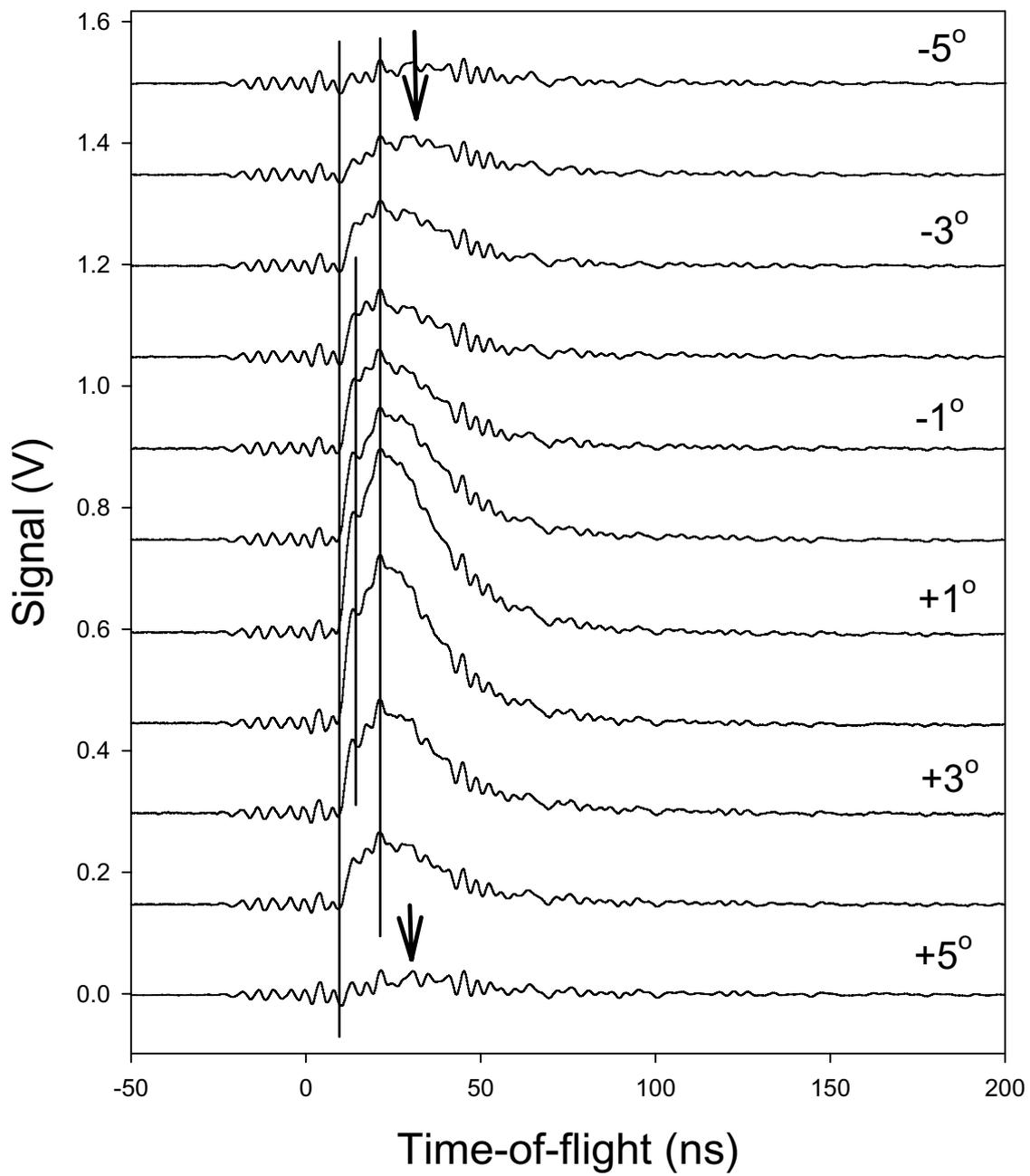

Fig. 5

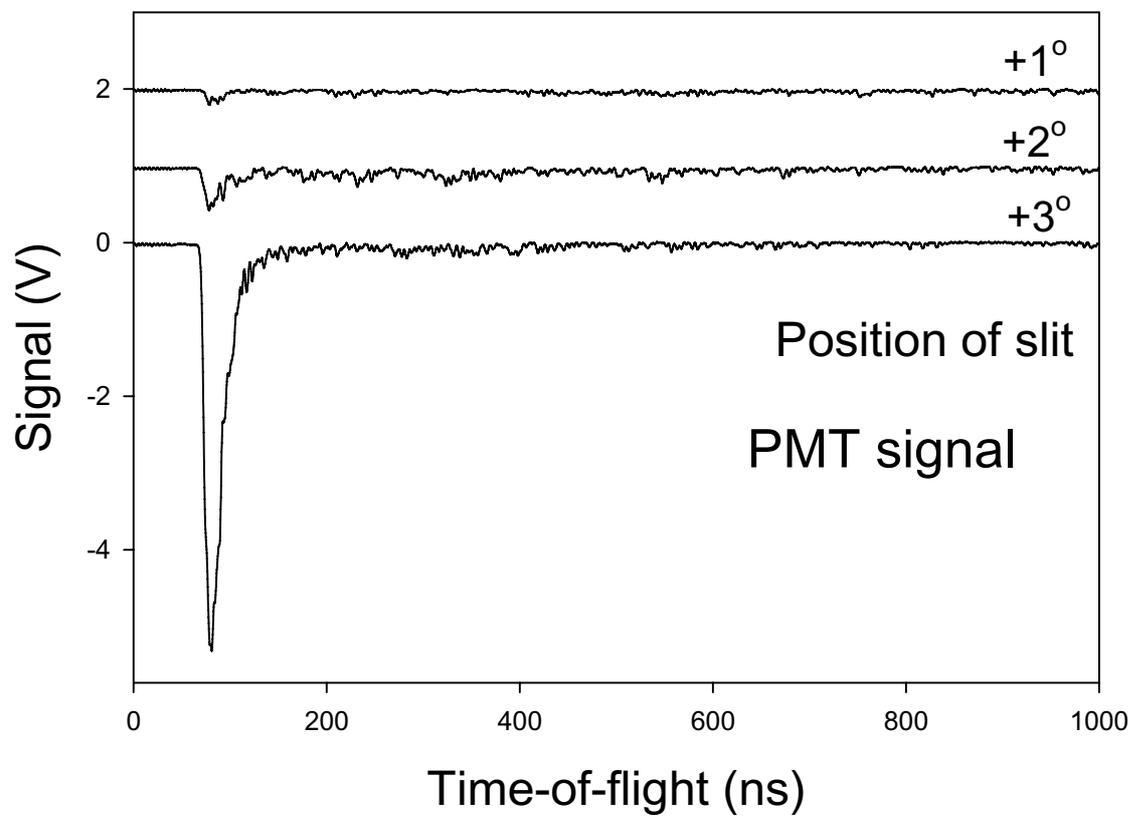

Fig. 6

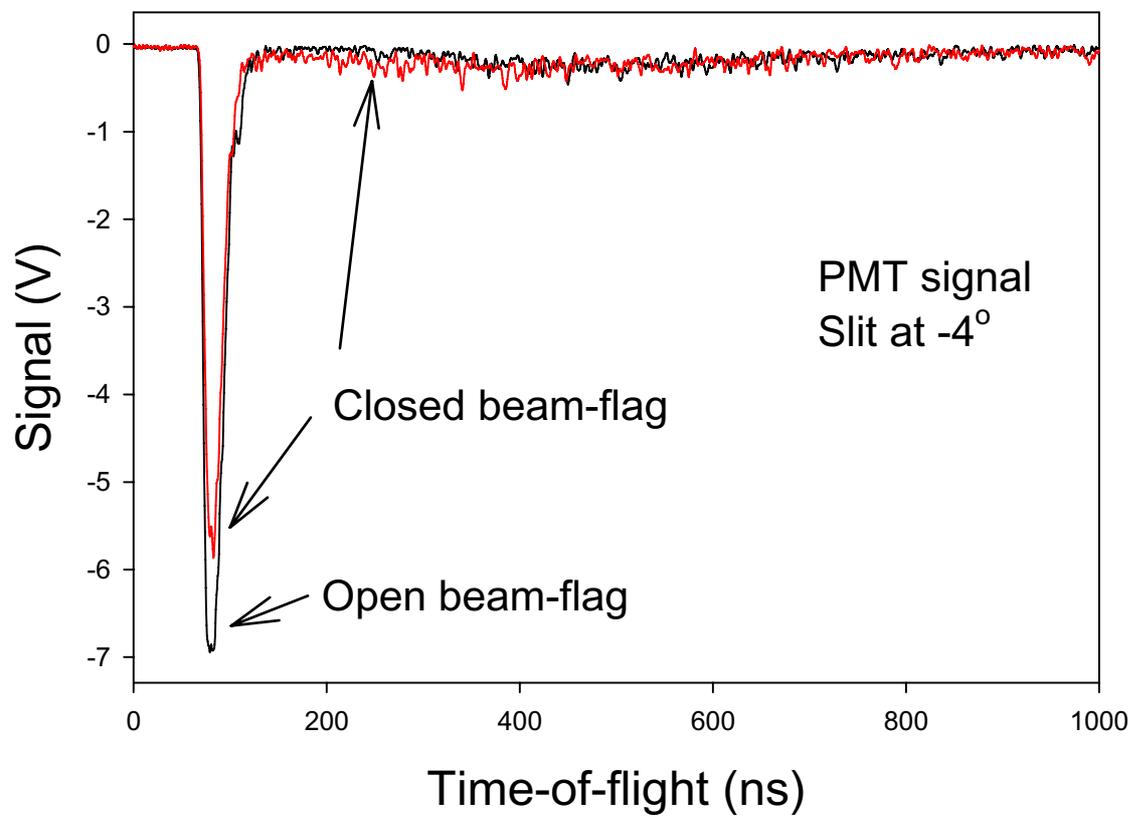

Fig. 7

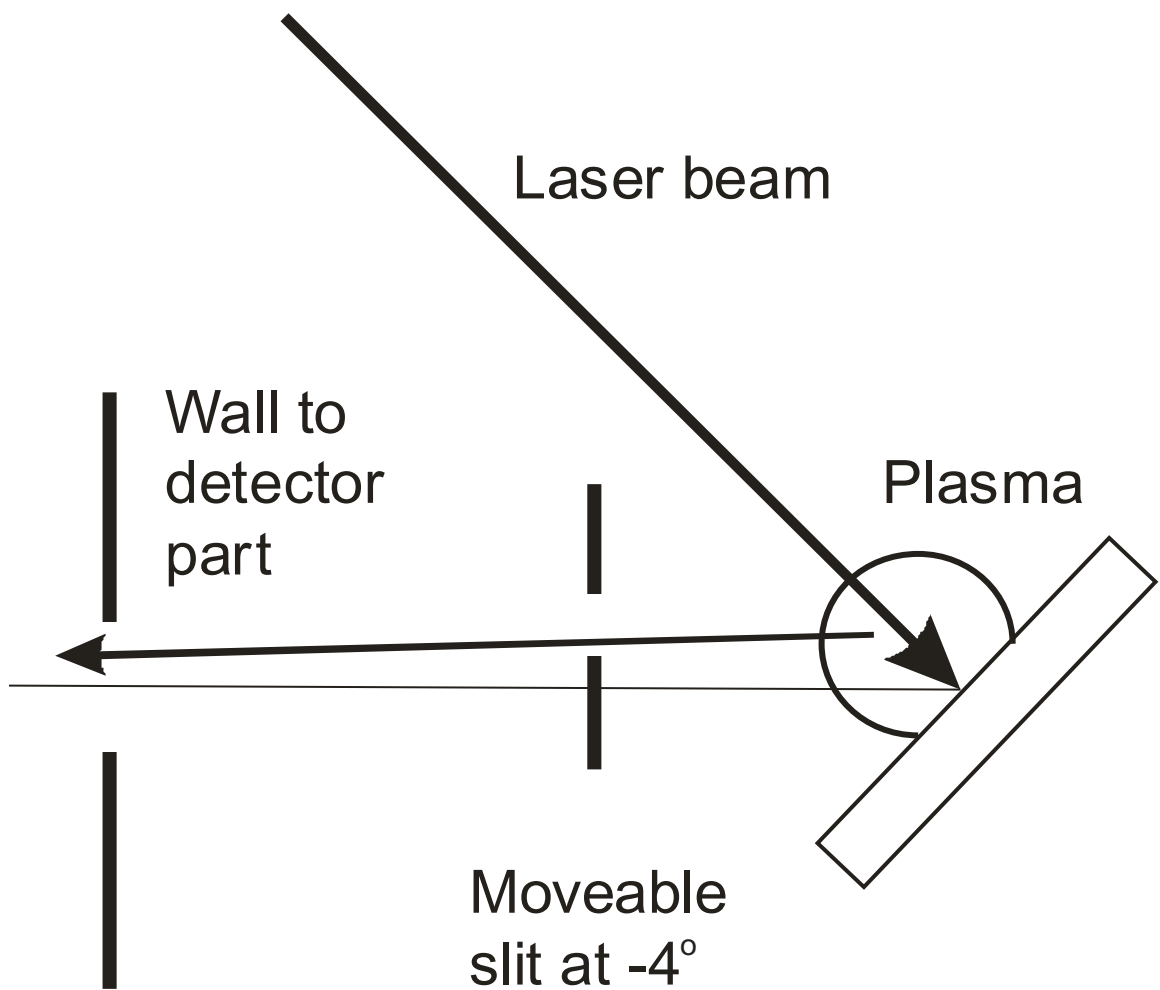

Fig. 8

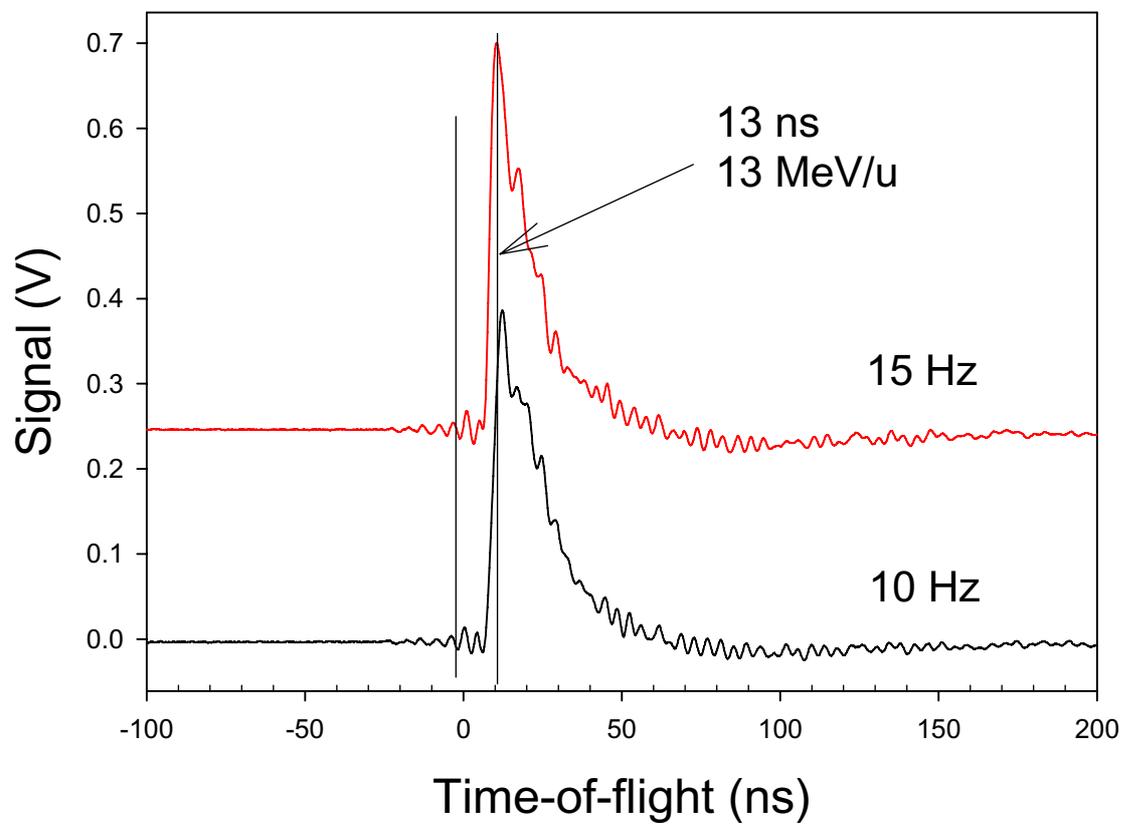

Fig. 9

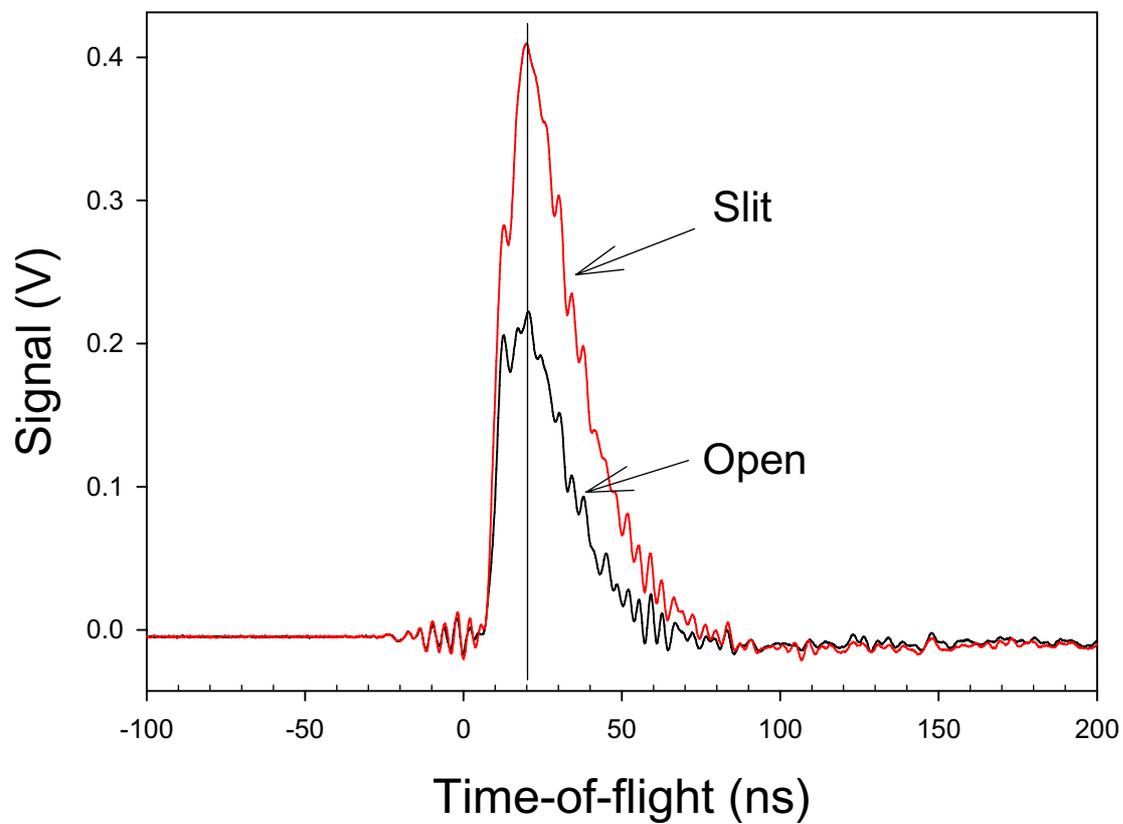

Fig. 10

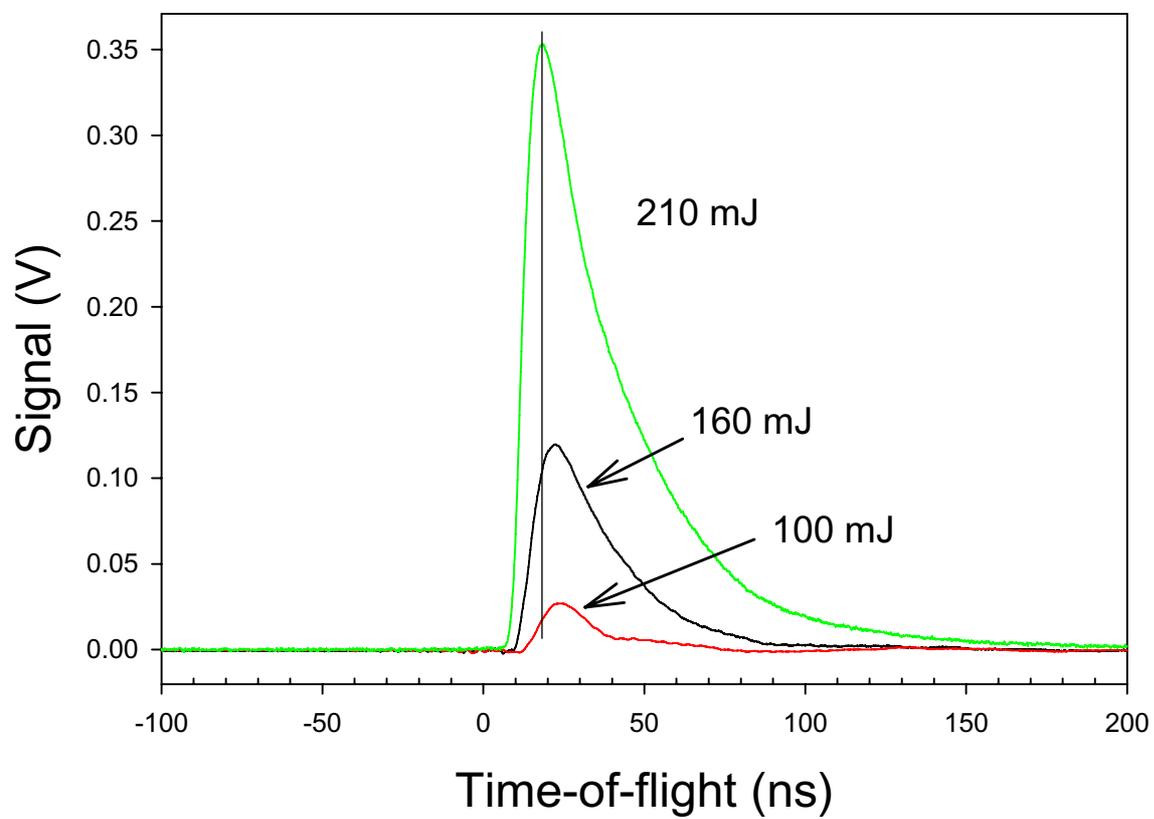

Fig. 11